\documentclass[a4paper]{article}

\usepackage{INTERSPEECH2021}
\usepackage{kotex}
\usepackage{amssymb}
\usepackage{xcolor}
\usepackage{hyperref}
\hypersetup{ colorlinks=true, linkcolor=blue, filecolor=magenta, urlcolor=cyan, }

\newcommand\bg[1]{\textcolor{blue}{#1}}

\renewcommand\bg[1]{\textcolor{black}{#1}}

\title{Broadcasted Residual Learning for Efficient Keyword Spotting}
\name{Byeonggeun Kim$^*$\thanks{* equal contribution}, Simyung Chang$^*$, Jinkyu Lee, Dooyong Sung}

\address{
  Qualcomm AI Research${}^{\dagger}$, Qualcomm Korea YH, Seoul, Republic of Korea  \thanks{  ${}^{\dagger}$ Qualcomm AI Research is an initiative of Qualcomm Technologies, Inc.}}
\email{\{kbungkun, simychan, jinkyu, dooysung\}@qti.qualcomm.com}

\begin{document}

\maketitle

\begin{abstract}
Keyword spotting is an important research field because it plays a key role in device wake-up and user interaction on smart devices. However, it is challenging to minimize errors while operating efficiently in devices with limited resources such as mobile phones. We present a \textit{broadcasted residual learning} method to achieve high accuracy with small model size and computational load. Our method configures most of the residual functions as 1D temporal convolution while still allows 2D convolution together using a broadcasted-residual connection that expands temporal output to frequency-temporal dimension. This residual mapping enables the network to effectively represent useful audio features with much less computation than conventional convolutional neural networks. We also propose a novel network architecture, Broadcasting-residual network (BC-ResNet), based on broadcasted residual learning and describe how to scale up the model according to the target device's resources. BC-ResNets achieve state-of-the-art $98.0 \%$ and  $98.7 \%$ top-1 accuracy on Google speech command datasets v1 and v2, respectively, and consistently outperform previous approaches, using fewer computations and parameters. Code is available at \href{https://github.com/Qualcomm-AI-research/bcresnet}{https://github.com/Qualcomm-AI-research/bcresnet}.
\end{abstract}
\noindent\textbf{Index Terms}: keyword spotting, speech command recognition, deep neural network, efficient neural network, residual learning

\section{Introduction}
Designing efficient architecture is an important topic in neural speech processing.
In particular, for keyword spotting (KWS), which aims to detect a predefined keyword, network efficiency is essential because it is usually performed in edge devices while requiring low latency. Recent efficient CNNs \cite{mobilenet1, mobilenet2, shufflenet, efficientnet} are usually made up of repeated blocks of the same structure and are based on residual learning \cite{residual} and depthwise separable convolutions \cite{depthwise_conv}. This trend continues in CNN-based KWS approaches, and they use either 1D temporal or 2D frequency$\times$temporal convolutions with pros and cons. In the case of using temporal convolution \cite{tcresnet, tenet, matchbox}, they require less computing than 2D approaches. However, the convolution's internal biases, such as translation equivariance, cannot be obtained for the frequency dimension. On the other hand, the approaches based on 2D convolution still require more computations than 1D methods despite efficient designs like using depthwise separable convolution \cite{res15, ds-resnet}.

\begin{figure}[t]
  \centering
  \includegraphics[width=\columnwidth]{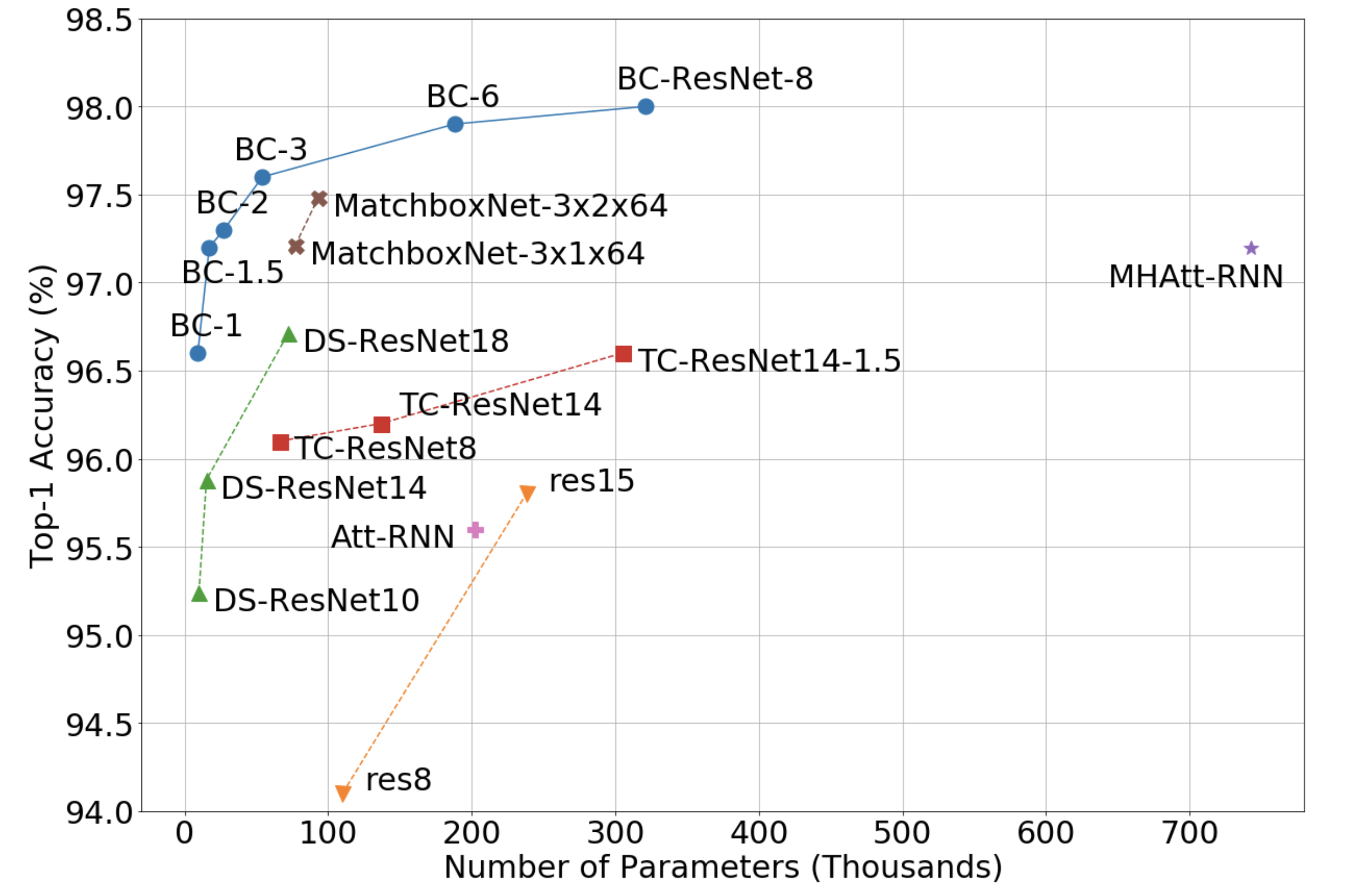}
  \vskip -0.05in
  \caption{\textbf{Model Size vs. Google speech command dataset v1 Test Accuracy.} The proposed BC-ResNets significantly outperform other KWS approaches. The smallest BC-ResNet-1 achieves $96.6 \%$ accruacy with less than 10k parameters. We scale the BC-ResNet-1 by channel width with a factor of 8, and BC-ResNet-8 achieves the state-of-the-art $98.0 \%$. The details are in Table~\ref{performance_table}.}
  \label{accxpar}
\end{figure}

In this paper, we introduce \textit{broadcasted residual learning} to address these problems of 1D or 2D convolution. Instead of processing all features in 1D or 2D, 
\bg{frequency-wise convolution performs on the 2D features.}
%2D convolution performs on the frequency dimension.
Then, we average the 2D features by frequency to get temporal features. After some temporal operations, we can apply residual mapping to the input 2D feature by broadcasting the 1D residual information. This learning method enables convolutional processing in the frequency direction to obtain the advantage of 2D CNNs while minimizing computational cost. Based on this residual learning method, we propose a novel network named broadcasting-residual network (BC-ResNet). BC-ResNet achieves top-1 accuracy 96.6\% and 96.9\%, respectively on the Google speech command datasets, v1 and v2 \cite{speechdataset} with less than 10k parameters. And by scaling up BC-ResNet, our method achieves state-of-the-art performance with a much smaller memory footprint than other keyword spotting methods, as shown in Figure~\ref{accxpar}.

Our contributions are summarized as follows:

(1) We introduce a brand new framework termed, \textit{broadcasted residual learning}, which utilizes the advantage of 1D temporal and 2D convolution while minimizing the increase of computation.

(2) We propose a novel model architecture, BC-ResNet, based on broadcasted residual learning and obtain a family of networks, BC-ResNets, by increasing the model width.

\begin{figure*}[t]
  \centering
  \includegraphics[width=1.\textwidth]{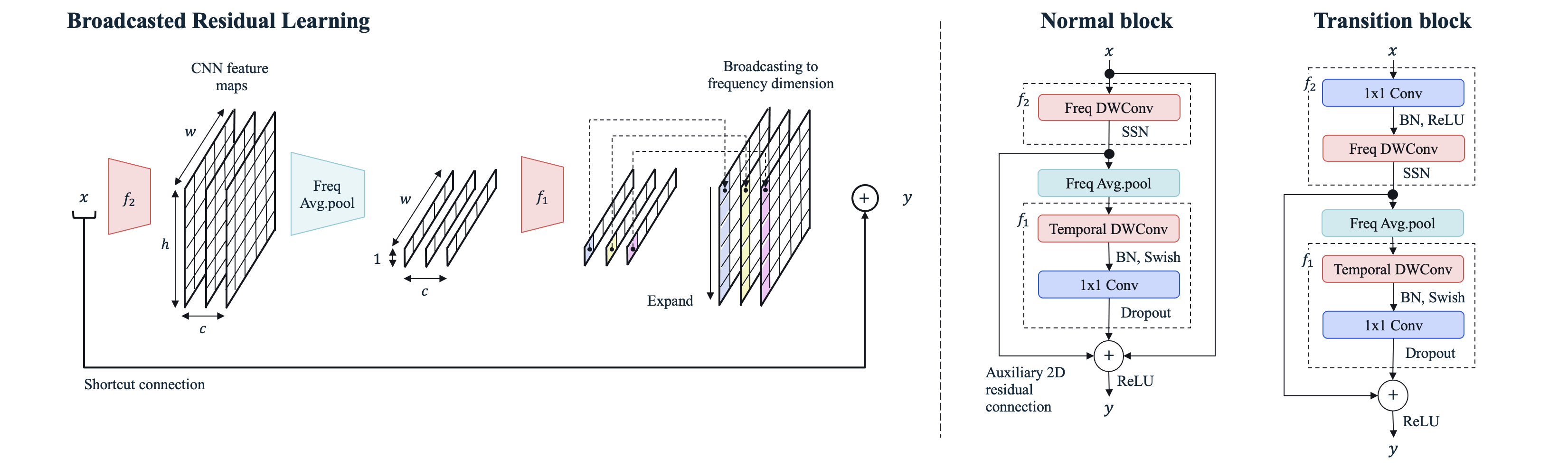}
  \vskip -0.15in
  \caption{\textbf{Left, Broadcasted Residual Learning} described in Equation~\ref{eq:broadcastsum1}, where $x\in \mathbb{R}^{c\times h \times w}$ with number of channels $c$. \textbf{Right, BC-ResBlock.} The BC-ResNet block contains a frequency-depthwise convolution with a SubSpectralNorm. Then the feature is averaged by frequency followed by temporal-depthwise separable convolution. 
%   with a Batchnorm, a swish activation and a dropout. 
  Temporal feature is broadcasted to 2D features at residual connection. In a transition block, we have an additional 1x1 convolution on the front to change the number of channel without identity shortcut.}
  \label{broadcastblock}
\end{figure*}

(3) The comprehensive experiments show our method's effectiveness in keyword spotting, and our model achieves state-of-the-art accuracy while reducing model parameters and computation.

\section{Proposed Method}

In this session, we introduce \textit{broadcasted residual learning} that can obtain the advantages of 1D and 2D convolution while minimizing an increase of computation. First, we define the broadcasted residual learning and describe a novel network architecture, the broadcasting-residual network (BC-ResNet). 
%, based on our learning method.
Following, we introduce a family of models, BC-ResNets, by channel scaling.

\subsection{Broadcasted Residual Learning}

A typical residual block \cite{residual} can be expressed as $y = x + f(x)$, where $x$ and $y$ are input and output features and function $f$ computes the residuals. Here the identity shortcut $x$ and residual $f(x)$ are \bg{usually} in same dimension and summed by simple addition. 
To utilize both 1D and 2D features together, we decompose the function $f$ into $f_1$ and $f_2$ which are the temporal and 2D operations, respectively. We average the 2D features after $f_2$ by frequency to get temporal features and expand the temporal feature back to the 2D shape after $f_1$. We repeatedly do the averaging and expanding at each residual block and propose \textit{Broadcasted residual learning}. Broadcasted residual learning employs a residual block of the form, 
\begin{align}\label{eq:broadcastsum1}
    & y = x + \textit{BC}(f_1(\textit{avgpool}(f_2(x)))),
\end{align}
as depicted in Figure~\ref{broadcastblock} left, where \textit{BC}, the \textit{Broadcasting} implies the expanding operation to frequency dimension, and \textit{avgpool} is average pooling by frequency dimension. In this way, broadcasted residual learning expands and adds a residual information to bigger dimension of identity shortcut.

\noindent \textbf{BC-ResNet Block} The overall architecture is depicted in Figure~\ref{broadcastblock} right. In equation~\ref{eq:broadcastsum1}, we ignore batch and channel dimensions for clarity, and the input feature $x$ is in $\mathbb{R}^{h\times w}$, where $h$ and $w$ correspond to the frequency and time dimensions, respectively. The 2D feature part, $f_2$ consists of a 3x1 frequency-depthwise convolution and SubSpectral Normalization (SSN) \cite{ssn} which splits the input frequency into multi-groups to separately normalize them. Here we use SSN instead of Batchnorm (BN) \cite{batchnorm} to achieve frequency-aware temporal features. After averaging by frequency, we get features in $\mathbb{R}^{1\times w}$. The $f_1$ is a composite of a 1x3 temporal depthwise convolution followed by BN, swish activation \cite{swish}, 1x1 pointwise convolution, and channel-wise dropout of dropout rate $p$. The broadcasting (BC) operation expands the feature in $\mathbb{R}^{1\times w}$ to $\mathbb{R}^{h\times w}$.

\begin{table}[t]
    \caption{\textbf{BC-ResNet-1.} Each row is a sequence of one or more identical modules repeated $n$ times with input shape of channel$\times$frequency$\times$time, total time steps $W$, and the number of output channels $c$. Changes in number of channels and downsampling by stride $s$ belong to the first block of each sequence of BC-ResBlocks. The temporal convolutions in all BC-ResBlocks use dilation of $d$.}
    \label{bc-resnet-0_structure}
    % \begin{center}
    % \begin{small}
    % \begin{sc}
    \centering
    \resizebox{\linewidth}{!}{
    \setlength{\tabcolsep}{1em}
    \begin{tabular}{c|c|c|c|c|c}
    \toprule
    Input 
    % \tiny{($F \times T \times C$)}
    & Operator & n & c & s & d  \\
    \midrule
    $1 \times 40 \times W$ & conv2d 5x5\scriptsize{ - BN - ReLU} & - & 16  & (2,1) & 1\\
    $16 \times 20 \times W$ & BC-ResBlock & 2 & 8 & 1 & 1 \\
    $8 \times 20 \times W$ & BC-ResBlock & 2 & 12 & (2,1) & (1,2) \\
    $12 \times 10 \times W$ & BC-ResBlock & 4 & 16 & (2,1) & (1,4)\\
    $16 \times 5 \times W$ & BC-ResBlock & 4 & 20 & 1 & (1,8) \\
    $20 \times 5 \times W$ & DWconv 5x5 & - & 20 & 1 & 1\\
    $20 \times 1 \times W$ & conv2d 1x1\scriptsize{ - BN - ReLU} & - & 32 & 1 & 1 \\
    $32 \times 1 \times W$ & avgpool & - & - & - & - \\
    $32 \times 1 \times 1$ & conv2d 1x1 & - & 12 & - & - \\
    \bottomrule
    \end{tabular}
    }
    % \end{sc}
    % \end{small}
    % \end{center}
    \vskip -0.1in
\end{table}

To be frequency convolution aware over the blocks, we add an \textit{auxiliary 2D residual connection} from 2D features. In summary, the proposed BC-ResBlock becomes
\begin{align}\label{eq:broadcastsum2}
    & y = x + f_2(x) + \textit{BC}( f_1(\textit{avgpool}(f_2(x)))).
\end{align}
We also define a \textit{transition block} (Where the number of In/Out channels are different) with two additional modifications; (a) add pointwise convolution where channel changing occurs followed by BN and ReLU activation. (b) No identity shortcut.

Using the proposed block, we can achieve an efficient KWS design while keeping 2D features. In a tiny network, pointwise convolution takes the most computations \cite{shufflenet}. We perform the temporal depthwise and the pointwise convolution on temporal features and reduce their computing load by a factor of $h$ compared to that of 2D depthwise separable convolutions.

\subsection{Network Architecture}

We design the base model, BC-ResNet-1, with parameters less than 10k as shown in Table~\ref{bc-resnet-0_structure}. The model has a 5x5 convolution on the front for downsampling by frequency with a BN and a non-linearity and followed by a total of 12 BC-ResBlocks. We split the blocks into four \textit{stages} which stand for a sequence of BC-ResBlocks whose activations are the same width. Inspired by \cite{efficientnet}, we explore the several choices and get the combination, 2, 2, 4, and 4 blocks for each stage, which implies that the model focuses more on performing high-level features. If the channel width $c$ is different from the input width, the first block of a stage is a transition block as in Figure~\ref{broadcastblock} right. To do residual learning, we keep the frequency and temporal dimension by using zero-padding for each depthwise convolution. After the BC-ResBlocks, there is a 5x5 depthwise convolution without zero-padding in frequency dimension followed by a pointwise convolution that increases the number of channels before average pooling. Here we add the 5x5 depthwise convolution to reduce the computations of the pointwise convolution behind it.

Many CNN-based KWS approaches use dilated convolutions to achieve required receptive fields \cite{res15, ds-resnet, matchbox}, and the proposed BC-ResNet also utilizes dilated convolutions. We empirically found that it is beneficial to keep the temporal dimension. Therefore we used stride $s$ in the frequency direction and dilation $d$ in the temporal dimension.

\noindent \textbf{Model Scaling} Previous KWS works usually scale their models by changing depth and width together \cite{tcresnet, matchbox, tenet} which makes it difficult to fit each computational or memory constraint. We explore compound \cite{efficientnet}, depth only, and width only scaling and decide to scale up the base model, BC-ResNet-1, by increasing the channel width $\tau$ times to get a BC-ResNet-$\tau$. Therefore the model is easy to scale with any predefined resources.

\section{Related Works}

\noindent \textbf{Efficient CNN-based KWS} While NASNet \cite{nasnet} and AmoebaNet \cite{amoeba} introduced automatic ways to optimize networks, there are successful handcrafted CNN designs such as MobileNets \cite{mobilenet1, mobilenet2} and ShuffleNet \cite{shufflenet} which utilize depthwise separable convolution, inverted bottleneck blocks, and channel shuffle. Inspired by the designs, there have been various CNN-based KWS approaches. \cite{res15} uses residual learning and DS-ResNet \cite{ds-resnet} adds depthwise separable convolutions upon \cite{res15}. TC-ResNet \cite{tcresnet} uses temporal convolution and treats frequency dimension as channel for better efficiency and TENet \cite{tenet} and MatchBoxNet \cite{matchbox} improve further with depthwise separable convolutions. While the approaches use either 1D or 2D convolution, the BC-ResNets suggest combining both.

\noindent \textbf{Other Approaches} There are other approaches that are not fully CNNs. Some of them use CNNs at the front and perform high-level features with recurrent neural networks (RNNs) \cite{att-rnn, orthogonalattn}. MHAtt-RNN \cite{mhatt-rnn} utilizes multi-head attention over it. On the other hand, there are automatic speech recognition based approaches \cite{LSTM-CTC,asrbase1}. Note that the approaches are successful but typically are not efficient in terms of the number of parameters compared to CNN-based approaches.

\section{Experiments}

\subsection{Experimental Setup}

\noindent \textbf{Datasets} We evaluate the performance of proposed BC-ResNets on Google speech commands datasets v1 and v2 \cite{speechdataset}. Version 1 contains 64,727 utterances from 1,881 speakers. There are total thirty words and we use ten classes of ``Yes'', ``No'', ``Up'', ``Down'', ``Left'', ``Right'', ``On'', ``Off'', ``Stop'', and ``Go'' with two additional classes ``Unknown Word (remaining twenty words)'' and ``Silence (no speech detected)'' following the settings of \cite{speechdataset}. Version 2 has 105,829 utterances from 2,618 speakers. There are 35 words and split into 12 classes as version 1. Each utterance is 1 sec long, and the sampling rate is 16 kHz. We divide the dataset into training, validation and testing set based on the validation and testing file lists \cite{speechdataset}. We re-balance the ``Unknown Word'' and ``Silence'' with the average number of utterances in the remaining ten classes following common settings of \cite{speechdataset, res15, mhatt-rnn} and especially we use the standard testing sets of v1 and v2 that the Google speech commands dataset offers. 
%After re-balancing dataset v1, we have 22,246, 3,093, 3,081 samples from training, validation and testing, respectively and 36,923, 4,445, 4,890 samples for that of dataset v2.

\noindent \textbf{Implementation Details} We use input features of 40-dimensional log Mel spectrograms with a 30ms window size and a 10ms frame shift. We followed the data augmentation settings of \cite{res15}, time shift in the range of -100 to 100ms, background noise \cite{speechdataset} with the probability of $0.8$, and SpecAugment \cite{specaugment} with two time and two frequency masks except time warping. We also found it beneficial to provide stronger augmentation as the model capacity increases. In specific, the smallest model, BC-ResNet-1 does not use SpecAugment and BC-ResNet-\{1.5, 2, 3, 6, 8\} use SpecAugment with \{1, 3, 5, 7, 7\} of frequency mask parameters, respectively with fixed temporal mask parameter of 20 \cite{specaugment}. \bg{Dropout rate is always $p=0.1$}. Also, we use SSN with five sub-bands \cite{ssn}. 
All models are trained for 200 epochs with stochastic gradient descent (SGD) optimizer with momentum to 0.9, weight decay to 0.001, minibatch size to 100, and a learning rate which linearly increases from zero to 0.1 over the first five epochs as warmup \cite{warmup} before decaying to zero with cosine annealing \cite{cosine_schedule}.

\begin{table}[t]
    \caption{\textbf{Impact of Broadcasted Residual Learning and Ablation Study.} We demonstrate how each component affects the base model, BC-ResNet-1, on Google speech command datasets v1 and v2. We show mean and standard deviation of Top-1 test accuracy (\%).  (averaged over 5 seeds).}
    \label{abalation_table}
    \centering
    \resizebox{\linewidth}{!}{
    \begin{tabular}{l|cccc}
    \toprule
    Model & v1 & v2 & \#Param & \#Mult \\
    \midrule
    ResNet-1D (w=2) & 95.0 $\pm$ 0.25 & 95.7 $\pm$ 0.46 & 27.3k & 3.2M\\
    ResNet-2D (w=1) & 94.8 $\pm$ 0.42 & 94.9 $\pm$ 0.32 & 7.9k& 5.9M \\
    ResNet-2D (w=1) w/ SSN & 95.5 $\pm$ 0.22 & 95.6 $\pm$ 0.24 & 9.4k & 5.9M \\
    BC-ResNet-Attn & 96.0 $\pm$ 0.14 & 96.2 $\pm$ 0.24 & 9.2k & 3.1M \\
    \midrule
    \textbf{BC-ResNet-1} & \textbf{96.6} $\pm$ \textbf{0.21} & \textbf{96.9} $\pm$ \textbf{0.30}  & \textbf{9.2k} & \textbf{3.1M} \\
    ~~ w/o auxiliary 2D residual & 96.2 $\pm$ 0.20 & 96.5 $\pm$ 0.10  & 9.2k & 3.1M \\
    ~~ w/o shortcut    & 96.4 $\pm$ 0.34 & 96.8 $\pm$ 0.18 & 9.2k & 3.1M \\
    ~~ w/o SSN & 96.1 $\pm$ 0.11 & 96.5 $\pm$ 0.12 & 7.8k & 3.1M \\
    ~~ w/o SSN (w=1.125) & 96.2 $\pm$ 0.26 & 96.7 $\pm$ 0.12 & 9.1k & 3.7M\\
    ~~ w/o 2D residual and SSN & 95.4 $\pm$ 0.29 & 95.7 $\pm$ 0.32 & 7.9k & 3.1M\\
    ~~ w/ Freq MaxPool & 96.3 $\pm$ 0.25 & 96.8 $\pm$ 0.11 & 9.2k & 3.1M\\
    \bottomrule
    \end{tabular}
    }
    \vskip -0.1in
\end{table}

\subsection{Impact of Broadcasted Residual Learning}
We compare BC-ResNet with fully 1D (ResNet-1D) and fully 2D (ResNet-2D) models to verify our method's effectiveness. These models consist of residual blocks with depthwise separable convolution instead of BC-ResBlock while maintaining the basic network architecture of BC-ResNet. ResNet-2D uses depthwise separable convolution with a 3x3 kernel, and ResNet-1D uses a 1x3 kernel. ResNet-1D requires a 1.1M multiply-accumulate (MAC) operation at the same width, which is smaller than BC-ResNet-1, so we scaled up the model by doubling its width for a fair comparison. Table 2 shows the comparison results of these baselines and our method.  

ResNet-1D has about three times more parameters, but it is still more than 1\% less accurate than our method. ResNet-2D has about 16\% fewer parameters than BC-ResNet by using batch normalization (BN) between depthwise convolution and pointwise convolution, and it requires a higher amount of computation due to more 2D operations. This 2D model shows about 2\% lower accuracy than our method. When we apply SSN to ResNet-2D instead of BN, ResNet-2D w/ SSN, the model size is similar to BC-ResNet-1, and it can obtain 0.7\% accuracy improvement without an increase in computation. This result shows that we can effectively apply SSN even in 2D CNNs. However, BC-ResNet still outperforms these baselines by a large margin. As shown in this result, broadcasted residual learning reduces the computation and can represent more discriminative information in keyword spotting.
BC-ResNet-Attn denotes the model using attention instead of broadcasted residual mapping like \cite{hu2018squeeze, lee2020urnet}. Sigmoid is applied to $f_1$ of equation~\ref{eq:broadcastsum2}, then element-wise multiplication is performed. It can be considered temporal-channel attention since the output of $f_1$ is a temporal feature. This model performs better than 1D and 2D baselines, but BC-ResNet is still more than 0.6\% accurate.

\begin{figure}[t]
  \centering
  \includegraphics[width=\columnwidth]{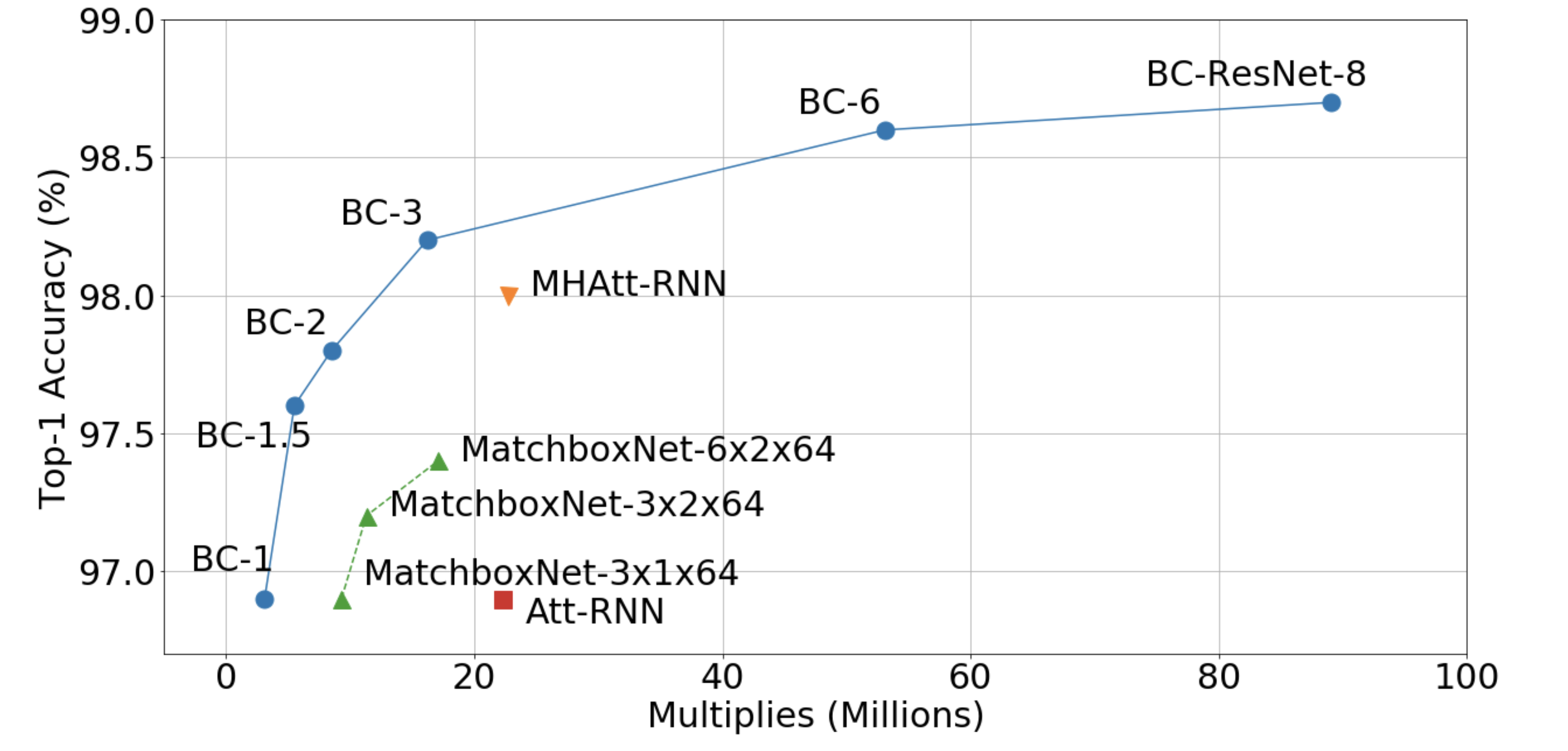}
  \vskip -0.05in
  \caption{\textbf{MACs vs. Google speech command dataset v2 Accuracy.}  Details are in Table~\ref{performance_table}.}
  \label{accxflop}
\end{figure}

\subsection{Ablation Study of BC-ResBlock}
Broadcasted residual learning plays a crucial role in BC-ResNet, but BC-ResBlock consists of %several
other core components; auxiliary 2D residual connection and SSN for frequency awareness of BC-ResBlocks.
% In this session, 
We evaluate how these components affect the model's performance.
BC-ResNet without the auxiliary 2D residual shows accurate results.
%with small parameters.
However, with
%due to
this additional connection, BC-ResNet can obtain about 0.4\% performance improvement without increasing the model size. And the shortcut connection of the identity also contributes slightly to performance.

`w/o SSN' shows the result when BN is used in BC-ResBlock instead of SSN, which eliminates inter-frequency deflection and provides frequency awareness. We can reduce some parameters by removing the SSN, but a more than 0.4\% accuracy drop occurs. We also compare the model that uses the base channel's size 9, `w/o SSN (w=1.125)', to compensate for the loss in parameters due to BN. This model requires about 20\% more computation but still has lower performance than BC-ResNet. \nocite{chang2018broadcasting} 
% These results show that SSN, which eliminates inter-frequency deflection and provides frequency awareness, also plays a key role in broadcasted residual learning. 
`w/o 2D residual and SSN' is the model without using both the 2D residual and SSN in BC-ResBlock. These two components help reduce information loss due to eliminating the frequency dimension. Removing all of them increases the error more significantly than when removing them one by one. These results show that the two components play an essential role in broadcasted residual learning.
We also evaluate the model,`w/ Freq MaxPool,' which uses max-pooling as a frequency dimension reduction method instead of average pooling. The model records a slightly lower accuracy than BC-ResNet-1. However, it still shows much higher accuracy than fully 1D and 2D models. It means that broadcasted residual learning can be effective with other frequency dimension reduction functions.

\begin{table}[t]
    \caption{\textbf{BC-ResNets Top-1 test accuracy (\%) on Google speech command datasets v1 and v2.} Each BC-ResNet is scaled up with a coefficient of $\tau$ in Section 4.3 and KWS approaches are grouped by accuracy for easier comparison. 
    % BC-ResNets match other approaches with much smaller parameters or multiplies.
    }
    \label{performance_table}
    \centering
    \resizebox{\linewidth}{!}{
    \setlength{\tabcolsep}{0.7em}
    \begin{tabular}{l|c|c|c|c}
    \toprule
    Model & v1 & v2 & \#Param & \#Mult \\
    \midrule
    Att-RNN \cite{att-rnn} & 95.6 & 96.9 & 202k & 22.3M \\ %multiplies
    ResNet-15 \cite{res15} & 95.8 & - & 238k & 894M \\ %multiplies
    DS-ResNet14 \cite{ds-resnet} &  95.9  & - & 15.2k & 15.7M \\
    TC-ResNet14-1.5 \cite{tcresnet} &  96.6  & -  & 305k & 6.7M \\ %multiplies % 13.4 M \\ FLOPs
    TENet12 \cite{tenet} & 96.6  & -  & 100k & 2.9M \\ %multiplies
    \textbf{BC-ResNet-1} & \textbf{96.6}  & \textbf{96.9}  & \textbf{9.2k}  & \textbf{3.1M} \\
    \midrule
    DS-ResNet18 \cite{ds-resnet} &  96.7  & - & 72 k  & 285 M \\
    \textbf{BC-ResNet-1.5} & \textbf{97.2} & \textbf{97.6}  & \textbf{17.2k} & \textbf{5.5M} \\
    \midrule
    MatchboxNet-3x1x64 \cite{matchbox} &  97.2  & 96.9  & 77k  & 9.3M \\
    Embedding+head \cite{embeddinghead} & - & 97.7  & 385k  & - \\
    \textbf{BC-ResNet-2} & \textbf{97.3} & \textbf{97.8}  & \textbf{27.3k}  & \textbf{8.5M} \\
    \midrule
    MatchboxNet-3x2x64 \cite{matchbox} &  97.5  & 97.2  & 93k  & 11.3M \\
    MatchboxNet-6x2x64 \cite{matchbox} &  - & 97.4  & 140k  & 17.1M \\
    MHAtt-RNN \cite{mhatt-rnn} &  97.2  & 98.0  & 743k  & 22.7M \\
    \textbf{BC-ResNet-3} & \textbf{97.6} & \textbf{98.2} & \textbf{54.2k}  & \textbf{16.2M} \\
    \midrule
    \textbf{BC-ResNet-6} & \textbf{97.9} & \textbf{98.6}  & \textbf{188k} & \textbf{53.1M} \\
    % \midrule
    \textbf{BC-ResNet-8} & \textbf{98.0} & \textbf{98.7}  & \textbf{321k}  & \textbf{89.1M} \\
    \bottomrule
    \end{tabular}
    }
    \vskip -0.1in
\end{table}

\subsection{Comparison with Baseline}

We compare the efficiency of KWS models in two aspects: performance per parameter and performance per MAC. In Figure~\ref{accxpar}, BC-ResNets are consistently efficient than other approaches in terms of accuracy per parameter on Google speech command dataset v1. We also compare MAC-accuracy curves on the datasets v2 as depicted in Figure~\ref{accxflop}. BC-ResNets achieve higher accuracy than 1D-based MatchboxNets \cite{matchbox} and state-of-the-art MHAtt-RNN \cite{mhatt-rnn} by a large margin while doing smaller computations. In the case of MatchboxNet, it shows better efficiency with dataset v1 compared to its v2 results as in Table~\ref{performance_table}, but still, BC-ResNets require x2.6 smaller number of parameters while achieving higher accuracy.

The details of the figures are in Table~\ref{performance_table}. We get the average performances of BC-ResNets in 10 random seeds each. The smallest model, BC-ResNet-1 matches the performance of 1D convolution-based approaches, TC-ResNet14-1.5 \cite{tcresnet} and TeNet12 \cite{tenet} with x10.9 smaller number of parameters while using 1D approach-level number of multiplies. BC-ResNet-3 works better than the state-of-the-art method, MHAtt-RNN \cite{mhatt-rnn} with x13.7 smaller number of parameters. The biggest model achieves the new state-of-the-art accuracy, 98.0 \% and 98.7 \% on Google speech command dataset v1 and v2 respectively, and is still x2.3 smaller than MHAtt-RNN \cite{mhatt-rnn}.

\section{Conclusions}

Existing CNN-based KWS approaches usually process all features by 1D or 2D convolutions with pros and cons. Using 1D convolution enables efficient design in terms of both the number of parameters and amount of computation, but it lacks the characteristics such as translation equivariance in the frequency direction. On the other hand, 2D convolution requires more computing compared to 1D approaches. To address the issues, we propose broadcasted residual learning that allows 1D and 2D features together. Broadcasted residual learning repeatedly averages 2D features to 1D features and expands 1D features back to the 2D. Leveraging the broadcasted residual learning and simple scaling by width, we design a family of networks called BC-ResNets and surpass state-of-the-art on Google speech command dataset v1 and v2.

\bibliographystyle{IEEEtran}

\bibliography{mybib}

\begin{thebibliography}{10}
\providecommand{\url}[1]{#1}
\csname url@samestyle\endcsname
\providecommand{\newblock}{\relax}
\providecommand{\bibinfo}[2]{#2}
\providecommand{\BIBentrySTDinterwordspacing}{\spaceskip=0pt\relax}
\providecommand{\BIBentryALTinterwordstretchfactor}{4}
\providecommand{\BIBentryALTinterwordspacing}{\spaceskip=\fontdimen2\font plus
\BIBentryALTinterwordstretchfactor\fontdimen3\font minus
  \fontdimen4\font\relax}
\providecommand{\BIBforeignlanguage}[2]{{%
\expandafter\ifx\csname l@#1\endcsname\relax
\typeout{** WARNING: IEEEtran.bst: No hyphenation pattern has been}%
\typeout{** loaded for the language `#1'. Using the pattern for}%
\typeout{** the default language instead.}%
\else
\language=\csname l@#1\endcsname
\fi
#2}}
\providecommand{\BIBdecl}{\relax}
\BIBdecl

\bibitem{mobilenet1}
A.~G. Howard, M.~Zhu, B.~Chen, D.~Kalenichenko, W.~Wang, T.~Weyand,
  M.~Andreetto, and H.~Adam, ``Mobilenets: Efficient convolutional neural
  networks for mobile vision applications,'' \emph{arXiv preprint
  arXiv:1704.04861}, 2017.

\bibitem{mobilenet2}
M.~Sandler, A.~G. Howard, M.~Zhu, A.~Zhmoginov, and L.~Chen, ``Mobilenetv2:
  Inverted residuals and linear bottlenecks,'' in \emph{{CVPR}}.\hskip 1em plus
  0.5em minus 0.4em\relax {IEEE} Computer Society, 2018, pp. 4510--4520.

\bibitem{shufflenet}
X.~Zhang, X.~Zhou, M.~Lin, and J.~Sun, ``Shufflenet: An extremely efficient
  convolutional neural network for mobile devices,'' in \emph{{CVPR}}.\hskip
  1em plus 0.5em minus 0.4em\relax {IEEE} Computer Society, 2018, pp.
  6848--6856.

\bibitem{efficientnet}
M.~Tan and Q.~V. Le, ``Efficientnet: Rethinking model scaling for convolutional
  neural networks,'' in \emph{{ICML}}, ser. Proceedings of Machine Learning
  Research, vol.~97.\hskip 1em plus 0.5em minus 0.4em\relax {PMLR}, 2019, pp.
  6105--6114.

\bibitem{residual}
K.~He, X.~Zhang, S.~Ren, and J.~Sun, ``Deep residual learning for image
  recognition,'' in \emph{{CVPR}}.\hskip 1em plus 0.5em minus 0.4em\relax
  {IEEE} Computer Society, 2016, pp. 770--778.

\bibitem{depthwise_conv}
F.~Chollet, ``Xception: Deep learning with depthwise separable convolutions,''
  in \emph{{CVPR}}.\hskip 1em plus 0.5em minus 0.4em\relax {IEEE} Computer
  Society, 2017, pp. 1800--1807.

\bibitem{tcresnet}
S.~Choi, S.~Seo, B.~Shin, H.~Byun, M.~Kersner, B.~Kim, D.~Kim, and S.~Ha,
  ``Temporal convolution for real-time keyword spotting on mobile devices,'' in
  \emph{{INTERSPEECH}}.\hskip 1em plus 0.5em minus 0.4em\relax {ISCA}, 2019,
  pp. 3372--3376.

\bibitem{tenet}
X.~Li, X.~Wei, and X.~Qin, ``Small-footprint keyword spotting with multi-scale
  temporal convolution,'' in \emph{{INTERSPEECH}}.\hskip 1em plus 0.5em minus
  0.4em\relax {ISCA}, 2020, pp. 1987--1991.

\bibitem{matchbox}
S.~Majumdar and B.~Ginsburg, ``Matchboxnet: 1d time-channel separable
  convolutional neural network architecture for speech commands recognition,''
  in \emph{{INTERSPEECH}}.\hskip 1em plus 0.5em minus 0.4em\relax {ISCA}, 2020,
  pp. 3356--3360.

\bibitem{res15}
R.~Tang and J.~Lin, ``Deep residual learning for small-footprint keyword
  spotting,'' in \emph{{ICASSP}}.\hskip 1em plus 0.5em minus 0.4em\relax
  {IEEE}, 2018, pp. 5484--5488.

\bibitem{ds-resnet}
M.~Xu and X.~Zhang, ``Depthwise separable convolutional resnet with
  squeeze-and-excitation blocks for small-footprint keyword spotting,'' in
  \emph{{INTERSPEECH}}.\hskip 1em plus 0.5em minus 0.4em\relax {ISCA}, 2020,
  pp. 2547--2551.

\bibitem{speechdataset}
P.~Warden, ``Speech commands: A dataset for limited-vocabulary speech
  recognition,'' \emph{arXiv preprint arXiv:1804.03209}, 2018.

\bibitem{ssn}
S.~Chang, H.~Park, J.~Cho, H.~Park, S.~Yun, and K.~Hwang, ``Subspectral
  normalization for neural audio data processing,'' in \emph{ICASSP 2021-2021
  IEEE International Conference on Acoustics, Speech and Signal Processing
  (ICASSP)}.\hskip 1em plus 0.5em minus 0.4em\relax IEEE, 2021, pp. 850--854.

\bibitem{batchnorm}
S.~Ioffe and C.~Szegedy, ``Batch normalization: Accelerating deep network
  training by reducing internal covariate shift,'' in \emph{{ICML}}, ser.
  {JMLR} Workshop and Conference Proceedings, vol.~37.\hskip 1em plus 0.5em
  minus 0.4em\relax JMLR.org, 2015, pp. 448--456.

\bibitem{swish}
P.~Ramachandran, B.~Zoph, and Q.~V. Le, ``Searching for activation functions,''
  in \emph{{ICLR} (Workshop)}.\hskip 1em plus 0.5em minus 0.4em\relax
  OpenReview.net, 2018.

\bibitem{nasnet}
B.~Zoph, V.~Vasudevan, J.~Shlens, and Q.~V. Le, ``Learning transferable
  architectures for scalable image recognition,'' in \emph{{CVPR}}.\hskip 1em
  plus 0.5em minus 0.4em\relax {IEEE} Computer Society, 2018, pp. 8697--8710.

\bibitem{amoeba}
E.~Real, A.~Aggarwal, Y.~Huang, and Q.~V. Le, ``Regularized evolution for image
  classifier architecture search,'' in \emph{{AAAI}}.\hskip 1em plus 0.5em
  minus 0.4em\relax {AAAI} Press, 2019, pp. 4780--4789.

\bibitem{att-rnn}
D.~C. de~Andrade, S.~Leo, M.~L. D.~S. Viana, and C.~Bernkopf, ``A neural
  attention model for speech command recognition,'' \emph{CoRR}, vol.
  abs/1808.08929, 2018.

\bibitem{orthogonalattn}
M.~Lee, J.~Lee, H.~J. Jang, B.~Kim, W.~Chang, and K.~Hwang, ``Orthogonality
  constrained multi-head attention for keyword spotting,'' in
  \emph{{ASRU}}.\hskip 1em plus 0.5em minus 0.4em\relax {IEEE}, 2019, pp.
  86--92.

\bibitem{mhatt-rnn}
O.~Rybakov, N.~Kononenko, N.~Subrahmanya, M.~Visontai, and S.~Laurenzo,
  ``Streaming keyword spotting on mobile devices,'' in
  \emph{{INTERSPEECH}}.\hskip 1em plus 0.5em minus 0.4em\relax {ISCA}, 2020,
  pp. 2277--2281.

\bibitem{LSTM-CTC}
Y.~Zhuang, X.~Chang, Y.~Qian, and K.~Yu, ``Unrestricted vocabulary keyword
  spotting using {LSTM-CTC},'' in \emph{{INTERSPEECH}}.\hskip 1em plus 0.5em
  minus 0.4em\relax {ISCA}, 2016, pp. 938--942.

\bibitem{asrbase1}
B.~Kim, M.~Lee, J.~Lee, Y.~Kim, and K.~Hwang, ``Query-by-example on-device
  keyword spotting,'' in \emph{{ASRU}}.\hskip 1em plus 0.5em minus 0.4em\relax
  {IEEE}, 2019, pp. 532--538.

\bibitem{specaugment}
D.~S. Park, W.~Chan, Y.~Zhang, C.~Chiu, B.~Zoph, E.~D. Cubuk, and Q.~V. Le,
  ``Specaugment: {A} simple data augmentation method for automatic speech
  recognition,'' in \emph{{INTERSPEECH}}.\hskip 1em plus 0.5em minus
  0.4em\relax {ISCA}, 2019, pp. 2613--2617.

\bibitem{warmup}
P.~Goyal, P.~Doll{\'a}r, R.~Girshick, P.~Noordhuis, L.~Wesolowski, A.~Kyrola,
  A.~Tulloch, Y.~Jia, and K.~He, ``Accurate, large minibatch sgd: Training
  imagenet in 1 hour,'' \emph{arXiv preprint arXiv:1706.02677}, 2017.

\bibitem{cosine_schedule}
I.~Loshchilov and F.~Hutter, ``{SGDR:} stochastic gradient descent with warm
  restarts,'' in \emph{{ICLR} (Poster)}.\hskip 1em plus 0.5em minus 0.4em\relax
  OpenReview.net, 2017.

\bibitem{hu2018squeeze}
J.~Hu, L.~Shen, and G.~Sun, ``Squeeze-and-excitation networks,'' in
  \emph{Proceedings of the IEEE conference on computer vision and pattern
  recognition}, 2018, pp. 7132--7141.

\bibitem{lee2020urnet}
S.~Lee, S.~Chang, and N.~Kwak, ``Urnet: User-resizable residual networks with
  conditional gating module,'' in \emph{Proceedings of the AAAI Conference on
  Artificial Intelligence}, vol.~34, no.~04, 2020, pp. 4569--4576.

\bibitem{chang2018broadcasting}
S.~Chang, J.~Yang, S.~Park, and N.~Kwak, ``Broadcasting convolutional network
  for visual relational reasoning,'' in \emph{Proceedings of the European
  Conference on Computer Vision (ECCV)}, 2018, pp. 754--769.

\bibitem{embeddinghead}
J.~Lin, K.~Kilgour, D.~Roblek, and M.~Sharifi, ``Training keyword spotters with
  limited and synthesized speech data,'' in \emph{{ICASSP}}.\hskip 1em plus
  0.5em minus 0.4em\relax {IEEE}, 2020, pp. 7474--7478.

\end{thebibliography}

\end{document}